\documentstyle[prl,aps,tabularx,psfig]{revtex}

\begin{document}

\title{Depth of maximum of extensive air showers
and cosmic ray composition above 10$^{17}$ eV in the
geometrical multichain model of nuclei interactions}
\draft
\author{T. Wibig}

\address{Experimental Physics Dept., University of \L odz,
Pomorska 149/153, 90-236 \L odz, Poland}

%\date{\today}

\maketitle

\begin{abstract}
The depth of maximum for extensive air showers
measured by Fly's Eye and Yakutsk experiments is analysed.
The analysis depends on the hadronic interaction model that
determine cascade development.
The novel feature found in the cascading process for nucleus--nucleus
collisions at high energies leads to a
fast increase of the inelasticity in heavy nuclei interactions
without changing the hadron--hadron interaction properties.
This effects the development of the extensive air
showers initiated by heavy primaries. The detailed calculations were
performed using the recently developed
geometrical multichain model and the CORSIKA simulation code.
The agreement with data
on average depth of shower maxima, the falling slope of the
maxima distribution, and these distribution widths are
found for the very heavy cosmic ray mass spectrum
(slightly heavier than expected in the diffusion model at about
$3\times 10^{17}$ eV and similar to the Fly's Eye composition at this
energy).
\end{abstract}


\section {Introduction}
When a very high energy cosmic ray particles enters the atmosphere
a large cascade of elementary particles is produced. As the cascade
develops, it grows until a maximum ``size'' (defined, for convenience,
as a number of charged particles, or electrons) is reached.
The amount of matter penetrated by the cascade when it reaches a maximum size
is denoted by $x_{\rm max}$ measured in g~cm$^{-2}$. Evaluating
$x_{\rm max}$ is a fundamental part of many cosmic ray studies, both
experimental and theoretical.

In general, the longitudinal development of an extensive air shower (EAS)
depends on primary particle energy and mass, and the nature of the
interactions of extremely high-energy particles. This last dependence
can be taken into account only by using a Monte Carlo technique to simulate
showers for fixed primary particle type and energy assuming a particular
mechanism responsible for secondary particles creation.

In this paper we would like to present results obtained using a
Monte Carlo technique for simulations of the longitudinal development
of EAS. The model we use is the geometrical multichain (GMC) model \cite{gmc}.
It is an extension to high and very high energies of
the two-chain model (G2C) described in Ref.~\cite{g2c}.
For cosmic ray purposes the GMC model, originally introduced
only for hadron--hadron collisions, was extended to describe nuclei interactions
in \cite{gmcinel}.

The detailed insight into the intranuclear collisions process gave
a surprising result which led to the revision of, e.g., the conventional
interpretation of connections between inelasticity in nuclei and hadronic
collisions and respective cross section ratios.
In the wounded nucleons picture of the high-energy nucleus--nucleus
interaction the non-zero time interval between excitation and hadronization
leads to the possibility which we will call hereafter second
step cascading:
the interaction of wounded nucleon from one (target or projectile) nucleus
with another nucleon from the same nucleus before hadronization occurs.

The great influence of the second step cascading process is straightforward.
For the
projectile (high laboratory energy) the excitation of one initially
untouched nucleus by a excited state going backward in anti-laboratory
(projectile) frame of reference
leads to the transfer a part of the nucleon energy
to the secondary produced particles. Their energy in the nucleus center of
mass system is rather small but after transformation to the laboratory system
the effect is expected to be considerable.

The second step cascading mechanism is introduced to the GMC event generator.
The quantitative description of its effect on
the longitudinal shower profile calculations and its consequences
are the main subject of this paper.

\section {Second step cascading and the GMC model
for nucleus--nucleus interactions.}

The idea of
wounded nucleons is generally accepted in high-energy nucleus--nucleus
interaction modelling. It can
be expressed as follows: once one of the projectile nucleons
interacts inelastically with the one from a target nucleus
intermediate states called ``wounded nucleons'' are
created. The ``spatial extension'' of the wounded nucleon is the same as
original nucleon before the collision. The subsequent
collisions inside the target nucleus take place before this
``excited state'' hadronize.
The non-zero time interval between excitation and hadronization
is essential for the interaction picture described in this paper.
Its value depends on particular definition (it is frame dependent)
but it is generally accepted to be of order of at least one fermi.

The $p$--$N$ collision acts in the wounded nucleon picture such that
each subsequent interaction of the once wounded nucleon leads to the
excitation of one of nucleons from the target. So, in a first
approximation, it adds to
overall average multiplicity about one half of $p$--$p$ multiplicity
(for respective interaction energy) and leads to the rather small decrease
of the incoming wounded nucleon energy (due to momentum and energy transfer).
Both features are confirmed by many experiments.

Similar results are obtained in dual-parton (DP)-like $p$--$N$
models were the
first interaction inside the target creates two quark-gluon strings stretched
generally between valence quarks of two colliding hadrons. The
next interactions each
produce two additional strings but their quark ends are formed
by the sea quarks on the projectile side thus leading to
$\sim$ $1/2$ of $p$--$p$ multiplicity addition and also to the small change
of the main projectile string energy and momentum.

The situation becomes much complicated when nuclei are involved
on both, target and projectile, sides. In this case there can also appear,
among the others, collisions of two wounded nucleons.
The introduction of wounded nucleons on both colliding
nuclei leads to the further decrease of the mean
multiplicity (and inelasticity) per one intranuclear interaction.
This can be solved quite naturally by careful investigation
of the so-called
intranuclear cascade. In both, the DPM- and Lund-like
interaction models the creation of all chains (strings) can be performed and
has been well known (e.g., in the FRITIOF realization of the Lund
picture \cite{fritiof} and DPMJETII for DPM \cite{dtujet} ) for some time.

The creation of wounded nucleons during the passage of one interacting nucleus
through the other is, in general, rather well established
\cite{capel}. The cascading of newly created hadrons inside
both nuclei is also quite natural. (The importance of this process
decreases
with the increase of the interaction energy due to the finite freeze-out time.)
Interactions between excited states (wounded nucleons, chains, strings) is
not so straightforward but has been extensively discussed (see, e.g.,
\cite{wnwn})
as important for quark-gluon plasma searches (and also
interpretations high transverse energy events). However the overall effect
on minimum bias event studies is not very significant \cite{capel}.
For the studies of the development of particle cascade in thick media
it is expected to be even less important
due to the fact that the energy of both incoming, earlier excited
states has to be conserved in the outgoing wounded states.

Second step cascading is defined as the process of
the interaction of wounded nucleon from one (target or projectile) nucleus
with another nucleon from the same nucleus before hadronization occurs.
A wounded nucleons excited to relatively high invariant mass in its nucleus
rest
frame of reference moves very fast due to the momentum and energy
conservation. The effect should enlarge with increasing
interaction energy.
The first reason for this is the increase of the wounded nucleon energy
in the nucleus rest frame. If the energy is high enough, the rise of the
nucleon--nucleon cross section start to play a role in an increase of
the probability of subsequent interaction. This probability also
increases due to the increase of the range of the wounded nucleon
within the nucleus (Lorentz dilatation of the freeze-out time).
If this time (or the Lorentz $\gamma$ factor) is high
enough, almost all wounded
nucleons will abandon the nucleus before secondary hadron creation starts.
Of course there
always could be cases when wounded nucleon hadronize inside nucleus,
e.g., single-diffraction-type excitations to any small invariant masses.

The second step cascading mechanism
as described above is based on a wounded nucleon
idea thus it should act
no matter what interaction model is applied for each nucleon--nucleon
interaction (DPM, Lund etc.).
The results presented in this paper are obtained with the GMC model.

The GMC model of the nucleon--nucleon (hadron--hadron)
collision differs from the others
(the DPM or Lund-type relativistic string model) in the
details of
the treatment of the creation of ``wounded nucleons''
(excited states, chains,
strings).
In the GMC model a phenomenological description of the chain creation is
used with very few parameters to be adjusted directly to the soft
hadronic interaction data, instead of using structure functions approach.
From one point of view,
this can be seen as a limitation of the model, but, on
the other hand, it can, of course, give a better data description. Some
connections between the GMC model and a structure function approach exist
anyhow, and both ways are in some sense equivalent.

The geometrization of the interaction picture is in the parametrization of
the multiparticle production process as a function of the impact parameter
of colliding hadrons.
A Detailed description of the model is given in \cite{g2c}.

During the initial state of the collision two intermediate objects,
called hereafter chains, are created. Their masses are determined by the
value of the impact parameter.
Next, each chain moves independently according
to energy--momentum conservation law for a time equal to
``freeze-out'' time (assumed to be a constant and energy independent
in a chain rest system).
Later the hadronization occurs.
Details can be found in Ref.~\cite{gmc} where some results are
given for
$\sqrt{s} \sim$ 20 GeV and SPS energies. In general, the particles are created
uniformly in the phase space with limited transverse momenta.
Flavours and spin states
are generated according to commonly accepted rules (with some slight
modifications). For very high energies (chain masses) the gluon
brehmsstrahlung process can take place, leading to prehadronization
breakups of the initial chain.

The wounded nucleon picture is adopted for GMC nucleus--nucleus interactions.
The detailed four-dimensional space-time history of each nucleon in both
colliding nuclei is traced.
Details of the nucleon distributions in nuclei are given in \cite{wsrap}.
It is obvious that two nucleons cannot be very close
to each other. In our calculations the minimal distance between
each two nucleons inside the nucleus was set equal to 0.8 fm.
Fermi motion is taken into account by Monte Carlo technique.

Each wounded nucleon in moving for the constant time measured in its
c.m.s. equal in the present calculation to 1 fm/$c$.
Calculation shows that this is enough for the wounded nucleon (in the most
of cases) to get out from its nucleus before the hadronization occurs,
if the interaction energy in the laboratory frame of about 1 TeV/nucleon
or more.
This means that for high energies, sometimes wounded nucleons excited to the
high masses collide with ordinary nucleons from its own nucleus. The GMC
procedure described in \cite{g2c}, without any changes, is used in such cases.
After hadronization, newly created hadrons are released from the
interaction without eventual re-cascading, which is quite reasonable for
the energy region of interest.

The remaining nucleons (spectators, both: from target and the projectile),
after collision form a fragment of nuclear matter which is, in most of
cases, highly excited (on the nuclear level). The excess of the energy can be
released by evaporation of protons, neutrons, or alphas, or by fragmentation
to
lighter nuclei. These very complicated nuclear processes can be, however,
described using simply formulae with good enough accuracy for our
purposes. In the GMC event generator we have used the method given in
\cite{jncfra}.

Among the many characteristics of any particular interaction model, the one
known as inelasticity
is found to be very closely connected with the longitudinal shower profile
(defined as the number of particles seen on a particular depth of the
atmosphere
(measured for convenience in g cm$^{-2}$)).
Generally, the inelasticity
can be understood as a fraction of the interaction energy transferred to the
secondary particles created in the interaction. The remaining energy
is carried by the ``leading particle'' and transported downward to
the next interaction in the hadronic cascade.

Knowledge of the inelasticity and its energy dependence of the
model used is important for a proper EAS data interpretation
(e.g. primary particle mass evaluation, as it will be discussed below).
On the other hand, experimental information about the
development of EAS can improve our knowledge about the physics
of the interaction processes not available in any other way.

The inelasticity can be defined in many ways.
In this paper we use the following definition:

\begin{eqnarray}
K_{\cal NN} =
{{\langle { \rm Energy\ carried\ by\ produced\ secondaries }\rangle} \over
{ \rm Initial\ energy\ of\ the\ projectile\ nucleus}} =
\ \ \ \ \ \ \ \ \ \ \ \ \ \ \ \ \ \ \ \ \ \ \ \
\nonumber \\
 { E_{\rm in}  -  E_{\rm nuclei}  -  \left( E_{\rm proton} + E_{\rm neutron}
 -  E_{\rm antiproton} -  E_{\rm antineutron}\right)  \over
{E_{\rm in}}} ,
\label{knn}
\end{eqnarray}

\noindent
where $E_{\rm in}$ is the incoming particle energy, $E_{\rm nuclei}$ is the
energy of all nuclei remaining from colliding
nuclei (if there are any), and $E_{\rm proton, neutron,...etc.}$ are energies
of all protons, neutrons,... etc. outgoing from the interaction.
All energies are given in laboratory system of reference.
In \cite{gmcinel} inelasticities were calculated using a little different,
approximative method. All the results presented in this paper are obtained
with the help of the GMC Monte Carlo event generator. Differences between
these two approaches are not significant and they result from many
constrains existing in real simulated events but not included in
the average and
simplified picture discussed in \cite{gmcinel}. All the conclusions
from \cite{gmcinel} are confirmed by detailed Monte Carlo calculations.
The GMC model
inelasticity for $p$--air is presented in Fig.~\ref{kfenppa} by a thick
solid curve. (For the comparison inelasticities of the other models
discussed below are also presented.)
As one can see, it is almost constant over a wide range of
interaction energies.

\begin{figure}
 \centerline{
 \psfig{file=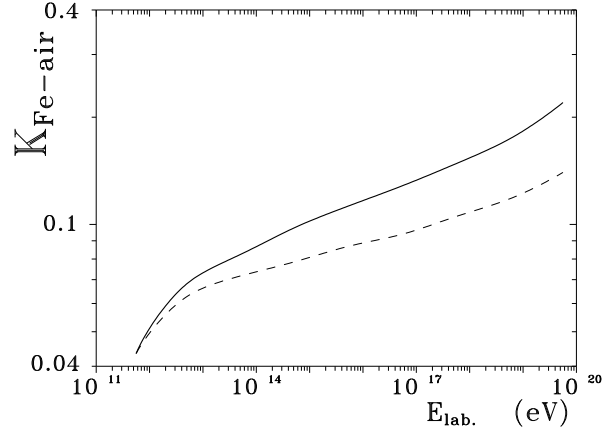,width=9.5cm}}
\caption{Inelasticity coefficient $Fe$--$air$ collisions as a function of
laboratory energy of incoming iron nucleus.
The solid and dashed lines were obtained with and without second step cascading
process taken into account.}
\label{kfenppb}
\end{figure}

In this paper we mainly want to discuss the role of the second step cascading
process in EAS development. The quantitative evidence for the change in
inelasticity between the model with and without this
process is given in Fig.~\ref{kfenppb},
where the inelasticities for iron--air
interactions are presented.

It should be mentioned that the
role of second step cascading decrease with decreasing
number of nucleons in the projectile, and for the proton interaction
on nuclear targets it (almost) vanishes. A primary wounded proton has
no possibility of finding an unwounded nucleon in the projectile, and
on the target side the amount of laboratory energy transferred during
the second step cascading is very small.

\section{Data and interpretation}
\subsection{Original interpretations of
Fly's Eye
and Yakutsk shower development results interpretation}

The Fly's Eye detectors consist of 103 mirrors of 1.5 m diameter
The details are given in Ref.\cite{FEexper}. Showers are detected using the
fluorescence light emitted by nitrogen molecules excited by charged
particles. Unlike the more intense Cerenkov light, the scintillation light is
emitted isotropically, allowing detection of the shower development profile
in an almost model-independent way.

Fly's Eye data have been published successively.
In 1990, data of stereo Fly's Eye experiment were presented in \cite{cass90}.
About 1000 events of energies above 3$\times $10$^{17}$ eV were analysed. The
average value of $x_{\rm max}$ reported is 690$\pm$3$\pm$20 g~cm$^{-2}$
where the first error is purely statistical while the second one was originally
described as a maximal systematic shift of average $x_{\rm max}$ value
related mainly to the treatment of Cerenkov light contamination.
This result was compared with Monte Carlo simulations. They were
performed for iron and proton primaries and obtained $x_{\rm max}$
values are 803 and 705 g~cm$^{-2}$, respectively. The model of hadronic
interactions used is described very briefly and it seems to be a version of
rather conventional (without any extraordinary assumptions in extrapolations
of the physics known from lower energies) and well-justified
minijet model developed in further works of Fly's Eye group (by
Gaisser and Stanev) which are briefly discussed below.

Comparison of three numbers given above led to rather astonishing
conclusions.
Cosmic ray composition seems to be (at least) iron dominated, even
assuming that experimental data are biased toward the ``iron side''
by the largest plausible shift.

Another result presented in that paper concerns the primary energy dependence
of the $x_{\rm max}$ value. The elongation rate defined as the change of the
average depth of the shower maximum per decade of primary particle energy
is the parameter very widely used in EAS study. Fly's Eye value of between
10$^{16}$ and 6$\times $10$^{18}$ eV is equal to 69.4$\pm$5.0 g~cm$^{-2}$
(with the
event reconstruction systematic error approximated as +5 g~cm$^{-2}$).
Data show no significant deviation from the constant elongation rate in the
whole observed energy region. No comparison with simulations is given
at this point.

Enriched Fly's Eye results were given in 1993 \cite{GS93}. The sample of 2529
events of above 3$\times $10$^{17}$ eV collected during 2649 h was analysed
from the
point of view of EAS longitudinal development. That paper contains also
extensive description of Monte Carlo simulation procedures used.
Three representative models of high-energy interactions, which relate the
inelasticity to the energy in quite different ways, were chosen
and examined. The first is the ``statistical'' one. It is characterized
by a power-law rise of mean multiplicity (as a function of interaction energy)
and {\em decrease} of the inelasticity coefficient reproduced in
Fig.~\ref{kfenppa} by the short dashed curve. It reaches the value of
0.4 for proton--air interaction at extremely high energies
($\sqrt{s} \sim 10^5$ GeV).
The second model is the opposite case. The Kopeliovich-Nikolaev-Potashnikova
model (KNP) \cite{knp} is a rather extreme version of the general DP
(pomeron)
interaction picture. A large amount of incoming energy is
transferred to secondary particle creation. Inelasticity in the KNP
model reaches the value of 0.8 at $\sqrt{s} \sim 10^5$ GeV.
The third, minijet model, gives almost constant proton--air
inelasticity of about 0.6.

The interaction event generation was considerably
simplified
for extremely large EAS simulations.
The overall laboratory inelasticity behavior was the main,
if not the only, difference between all three models.
The shower particles were followed in simulations directly down to
1/300 of the primary energy per nucleon. Below this energy parametrizations
(based on low-energy Monte Carlo shower calculations) were used.
Such a treatment introduces, as was mentioned by the authors,
the systematic bias of about 10 g~cm$^{-2}$ of longitudinal shower
development.

The Fly's Eye reconstruction procedures and trigger effects were
also extensively examined. The conclusion of the previous paper about the
maximal systematic shift (not larger than 20 g~cm$^{-2}$) was confirmed.
After a very detailed discussion the authors concluded that ``pure
Monte Carlo'' development curves should be shifted by 25 g~cm$^{-2}$
deeper in the atmosphere. It should be remembered that the finally
presented \cite{GS93} calculation results were in fact corrected.
This shift allowed reinterpretation of the Fly's Eye $x_{\rm max}$ data.
The primary cosmic ray particles do not necessarily have to be ``heavier
than iron''. The confusion mentioned in the previous paper seemed to vanish.

While agreement with the $x_{\rm max}$ value was achieved, the
elongation rate measured was in strong disagreement with the
calculation results assuming constant cosmic ray mass composition.
The measured value of 75.3$\pm$4 g~cm$^{-2}$ is much higher than
the 49$\pm$3 g~cm$^{-2}$ for minijet and KNP models. (The statistical model
did not match the experimental data producing very late developing
showers.)

The solution is, of course, the decrese of primary cosmic ray
particle mass with increasing energy.
Between $4 \times 10^{17}$ and 10$^{18}$ eV
the iron fraction in the two-component (iron and protons only) mass
spectrum changes from 80\% to 56\% for KNP and from 80\% to 60\%
for the minijet model.

\begin{figure}
 \centerline{\hspace{1.5cm}
 \psfig{file=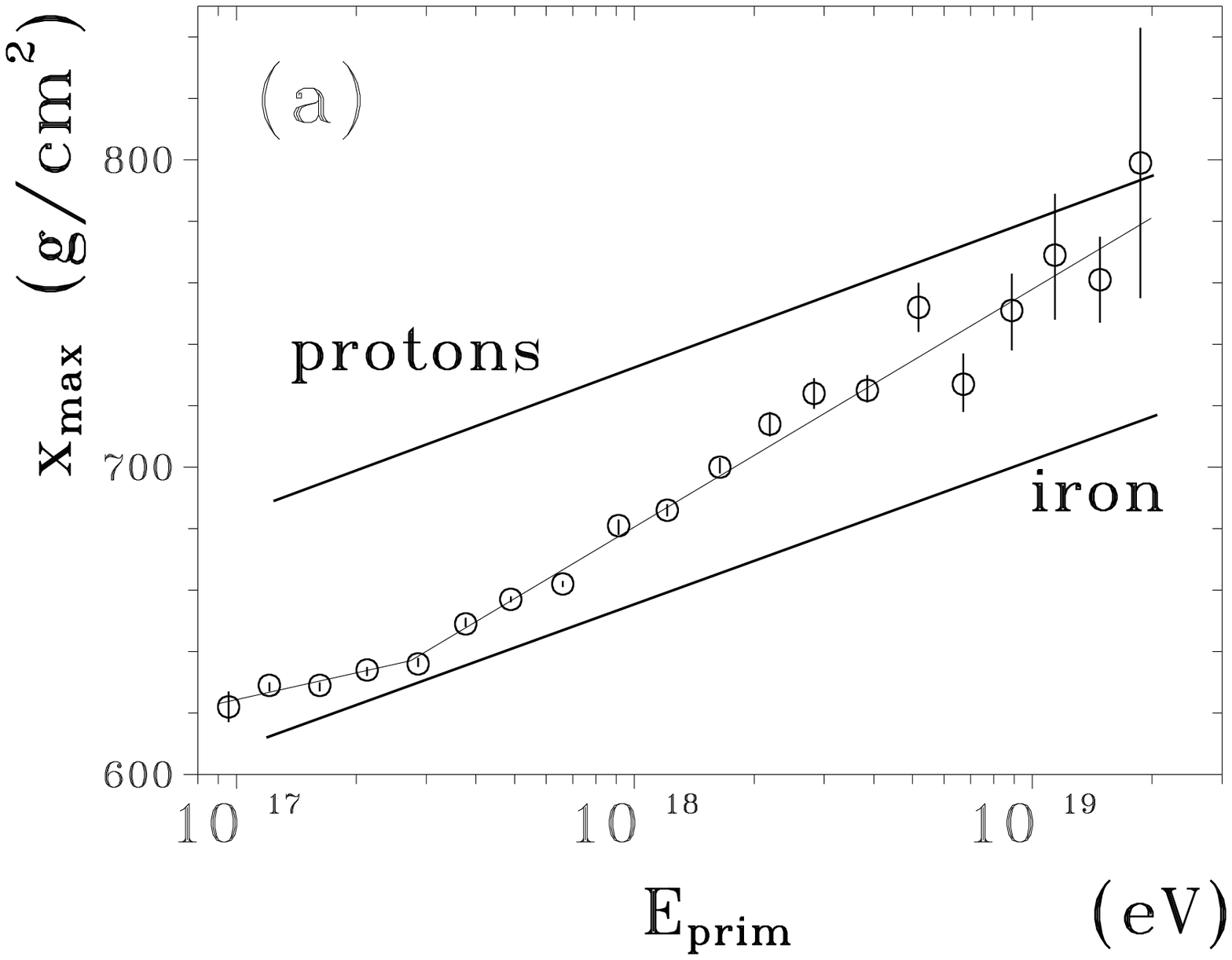,width=9.5cm}
 \psfig{file=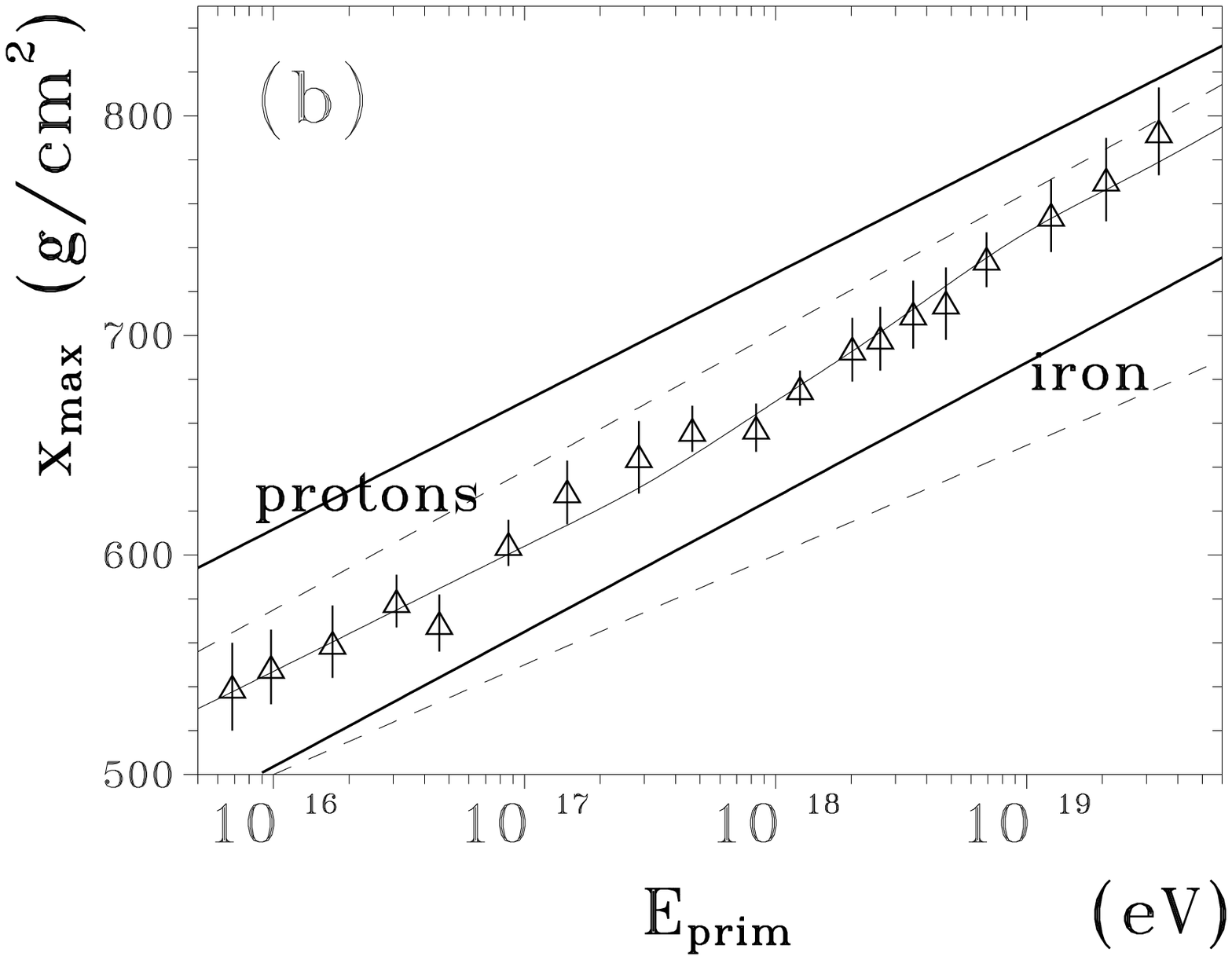,width=9.5cm}}
\caption{ Energy dependence of shower maximum
depth on primary particle energy.
Fly's Eye data points are taken as well as theoretical
predictions from [14] (a). Yakutsk data and dashed
lines are from Ref.[16], solid lines from Ref.[17] (b).}
\label{xmaxexp}
\end{figure}

In \cite{bird93} the 5477 event sample of energies above
4$\times $10$^{17}$ eV
was analysed and the final conclusion of \cite{GS93} was confirmed.
The famous figure showing the $x_{\rm max}$ as a function of primary
shower energy was given. It has been reproduced later in many papers
(see, e.g., \cite{bird94}).
It is again presented in Fig~\ref{xmaxexp}($a$). The two straight lines
represent results of KNP model calculations (with 25 g~cm$^{-2}$ shift
as discussed above) for primary protons and iron nuclei while the broken line
is obtained with two component cosmic ray mass spectrum
where the iron fraction is

\begin{equation}
{{\rm iron \  flux}  \over {\rm proton \  flux}} \ = \
10.0^{\textstyle -\: 0.887\: [\log (E) \:  - \:  18.5]}  \ .
\end{equation}

The importance of the idea of two component spectrum is
connected with the interpretation of the other
spectacular Fly's Eye result concerning cosmic ray energy spectrum.
Cosmic ray flux below 10$^{18.5}$ eV is, according to this hypothesis,
of Galactic origin and dominated by heavy (iron) nuclei. Above
this energy the extragalactic component, dominated by proton, is seen.
Both components have a simple power-law energy spectra but with
different slopes.

The agreement with all major Fly's Eye experimental results
(the energy spectrum, longitudinal development and absence of detectable
anisotropy of cosmic ray arrival directions) is remarkable.

Another giant EAS array which allows for the observation of shower
longitudinal profile is the Yakutsk Cerenkov experiment. The results
concerning $x_{\rm max}$ from \cite{diakoncalga}
are given in Fig.~\ref{xmaxexp}($b$)
together with some theoretical prediction from \cite{kal95}
(solid lines) and from \cite{diakoncalga} (dashed lines).
Again the straight lines represent ``pure'' proton and iron
cosmic ray mass spectra. The solid curve is described as a result for a mixed
composition and quark-gluon string model of high-energy interactions.
The composition is transformed in energy according to
Peters-Zatzepin diffusion model from
the normal composition before the knee of the primary energy
spectrum to one enriched by heavy nuclei at about 10$^{17}$ eV. Above
some critical energy of order of $3\times 10^{17} \: Z$ eV the diffusion
propagation stops leading to the restoration of the normal composition
above 10$^{19}$ eV.

The original interpretations of Yakutsk result are different than for
the Fly's Eye one.
Concerning theoretical predictions from \cite{kal95}, one can see that
the lightening of primary cosmic ray particle mass result is in a slightly
better agreement with data than the result for constant mass spectrum.
However, the statistical
significance of such conclusion is not great. The same,
but even less significant statement, can be given
using only the Yakutsk group calculations (dashed lines in
Fig.~\ref{xmaxexp}($b$)).

The problem is that the discrepancies between theoretical expectations
obtained by each group are extremely large.
For example, the depth of iron shower maximum
at 10$^{17}$ eV is 550 and 565 for \cite{diakoncalga} and
\cite{kal95} calculations, respectively.
(615 g~cm$^{-2}$ by \cite{bird93}!)
The source of these
discrepancies is mostly in the details of the interaction models assumed.
The model used in \cite{kal95} is a version of quark-gluon string
model with very fast increase of ``hard'' interaction cross section.
The comparison with the standard pomeron interaction picture
shows that the energy dissipation rate
in the EAS is faster than in the model with pomeron intercept
higher than 1.14.
Calculations in \cite{diakoncalga} are also based on quark-gluon string
model, but, as it can be seen in Fig.\ref{xmaxexp}($b$), the inelasticity
or and/or cross section has to increase with energy even faster.

The very fast shower development gives an agreement with more or less
``normal'' cosmic ray mass composition (corrected by the
diffusion model). However, the assumptions
about hadronic interaction at extremely high energies should be
treated with care.

All the models for extremely high-energy nuclei interactions used
in the papers discussed above are based on the
standard Glauber-type approach. This means that once the characteristics of
nucleon--nucleon interaction are fixed, the interactions between nuclei are,
in fact, determined. A change of interaction properties can only be made by
changing the hadronic interaction models.
To move the shower maxima
higher, the energy degradation has to be faster.
Two possibilities are apparent. The inelasticities or
the cross sections have to be increased.
But there are limits beyond which the
reality of a model becomes questionable or even vanishes.
Moreover, the proton--air cross section for energies of interests can only be
measured in the experiments like Fly's Eye or Yakutsk
by fitting the descend of the $x_{\rm max}$ distributions with
the exponential function of the form $\exp (\: -\: x_{\rm max}\: / \:
\lambda_{p-air} \: )$, where $\lambda_{p-air}$ is the mean proton interaction
length. The value of $\lambda_{p-air}$ measured by Fly's Eye experiment
\cite{GS93} is equal to 62.5$\pm$4 for energies of above 10$^{18}$ eV.
This left practically no room for meaningful changes.

\subsection{Longitudinal development of EAS calculations with the GMC model}

To compare high-energy cosmic ray data with some theoretical predictions
about the primary particle type or interaction model, Monte Carlo
programs have to be used to simulate the EAS development. In this work
we have used the CORSIKA code \cite{corsika} developed by the KASCADE
group.
Briefly, the code is the full four-dimensional simulation of the shower
development in the atmosphere. The electromagnetic component of the
shower is treated with the help of EGS \cite{egs} procedures. Except for the
Landau-Pomaranchuk effect, all the electromagnetic processes are included
in the program with an exactness sometimes exceeding the needed
for our purposes. The high-energy interaction models existing as options
in the distributed CORSIKA 5.20 package are: VENUS, QGSM, SIBYLL, HDPM,
and DPMJET.
They were
tested (also $x_{\rm max}$ predictions) in \cite{compa}.

The GMC model described above was included in this list by us. In this
paper only calculations with this model are presented.
To perform very high-energy shower simulations the method of ``thinning''
was used. It relate each particle in EAS to the ``weight'' which
allows one not to trace all of relatively low energy particles.
In the present calculations the cascade particles were followed
directly to 10$^{-4} --- 10^{-3}$
of primary particle energy per nucleon and below only
one of the secondary produced particle was chosen, but with
appropriate weight attached to it and to all their subsequent
interaction products.

\begin{figure}
 \centerline{\psfig{file=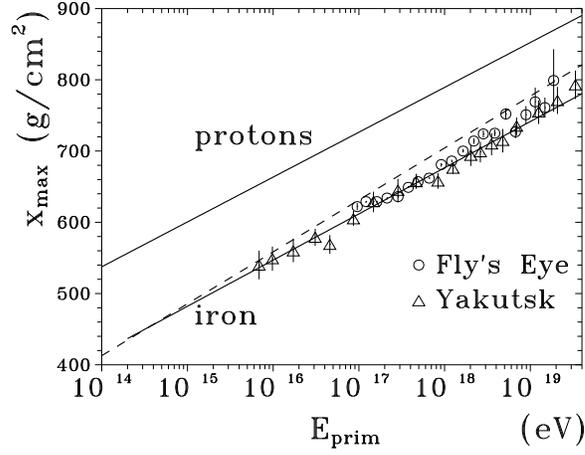,width=9.5cm}}
\caption{
Depth of the sower maxima calculated with the GMC model compared with the
Fly's Eye and Yakutsk data as a function of primary particle energy.
For the iron primaries the dashed line
represents the model without second step cascading process while
the solid one is obtained with this process taken into account.}
\label{xmax}
\end{figure}

\begin{figure}
 \centerline{
 \psfig{file=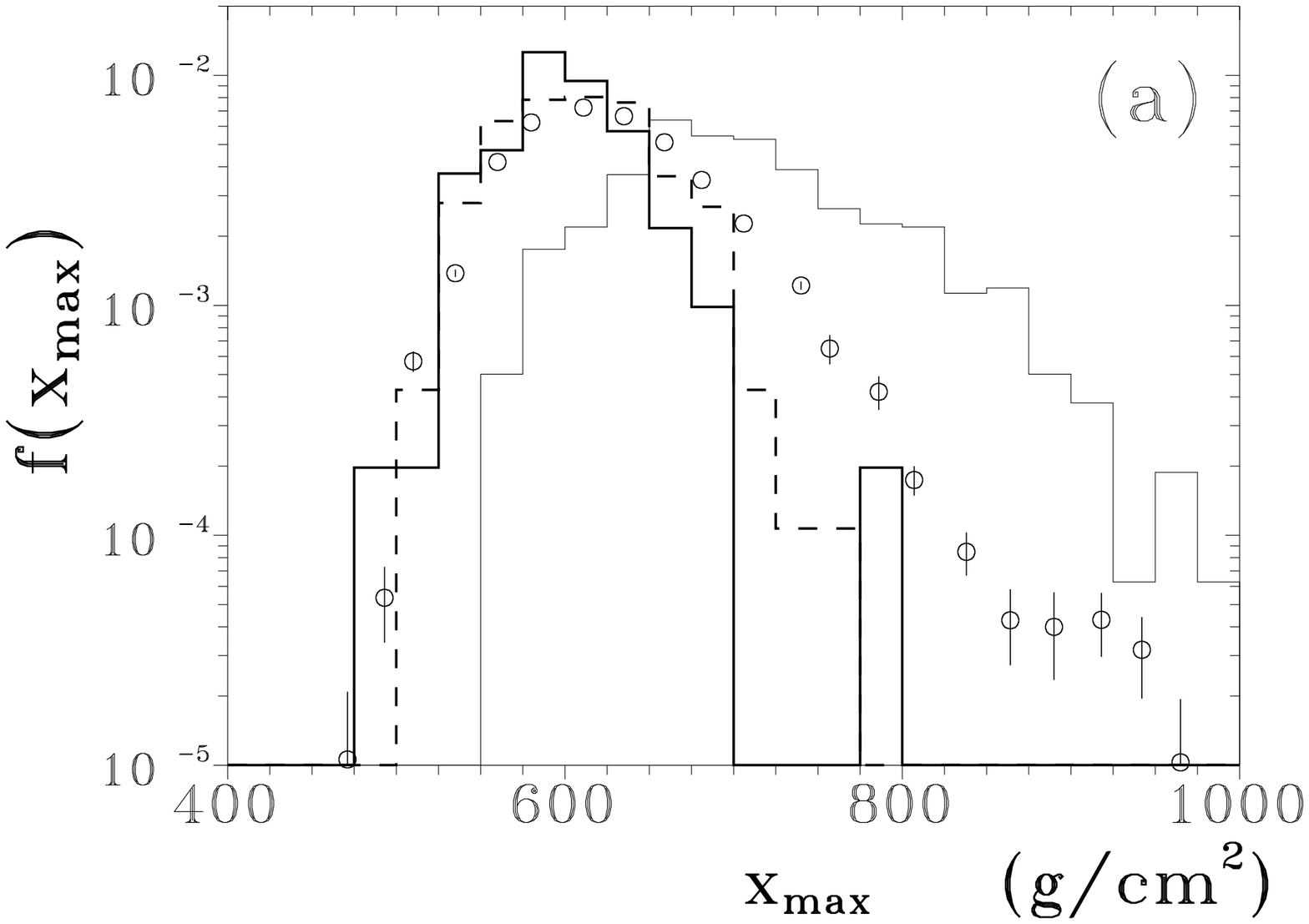,width=9.5cm}
 \psfig{file=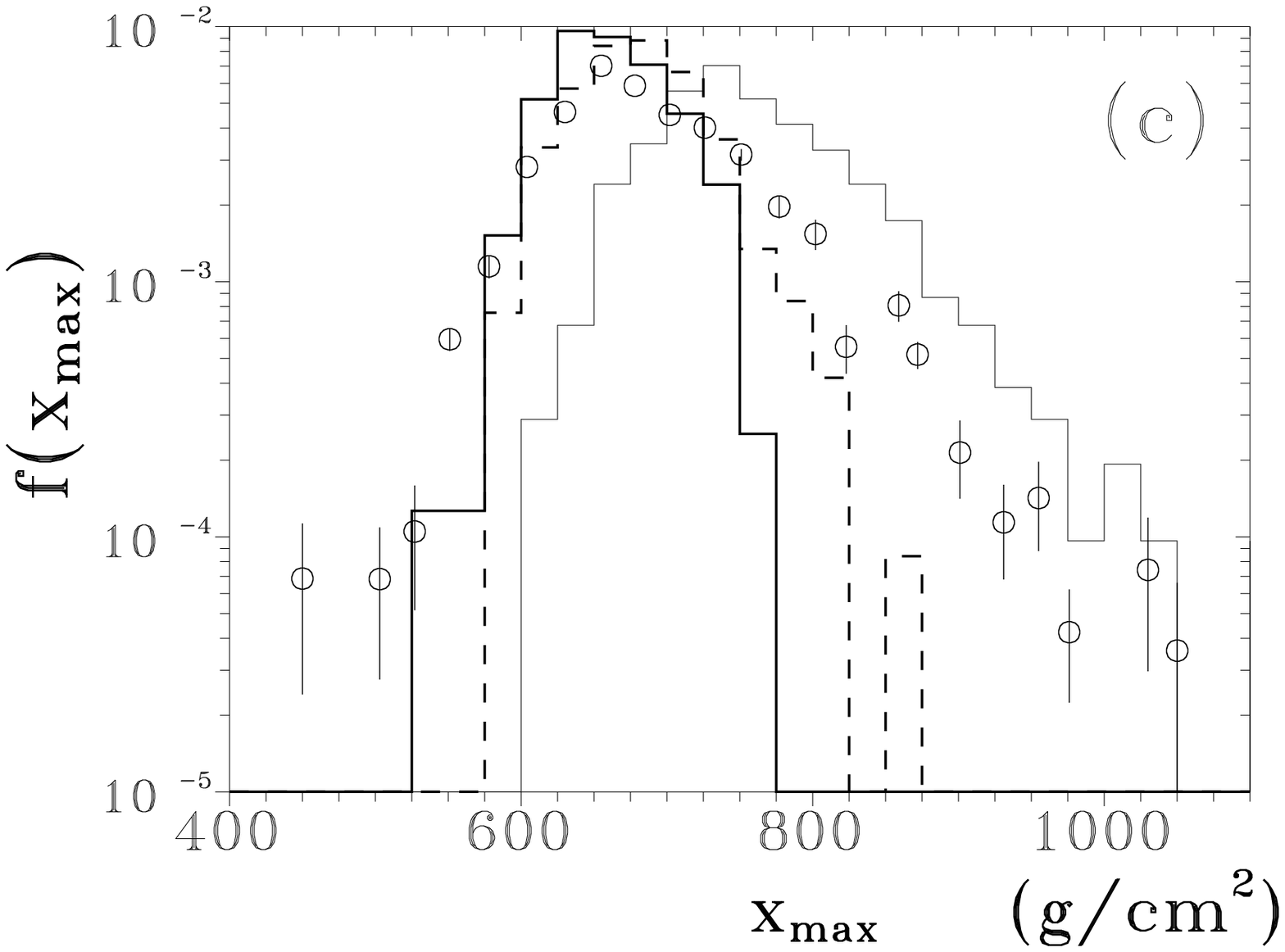,width=9.5cm}}
 \centerline{
 \psfig{file=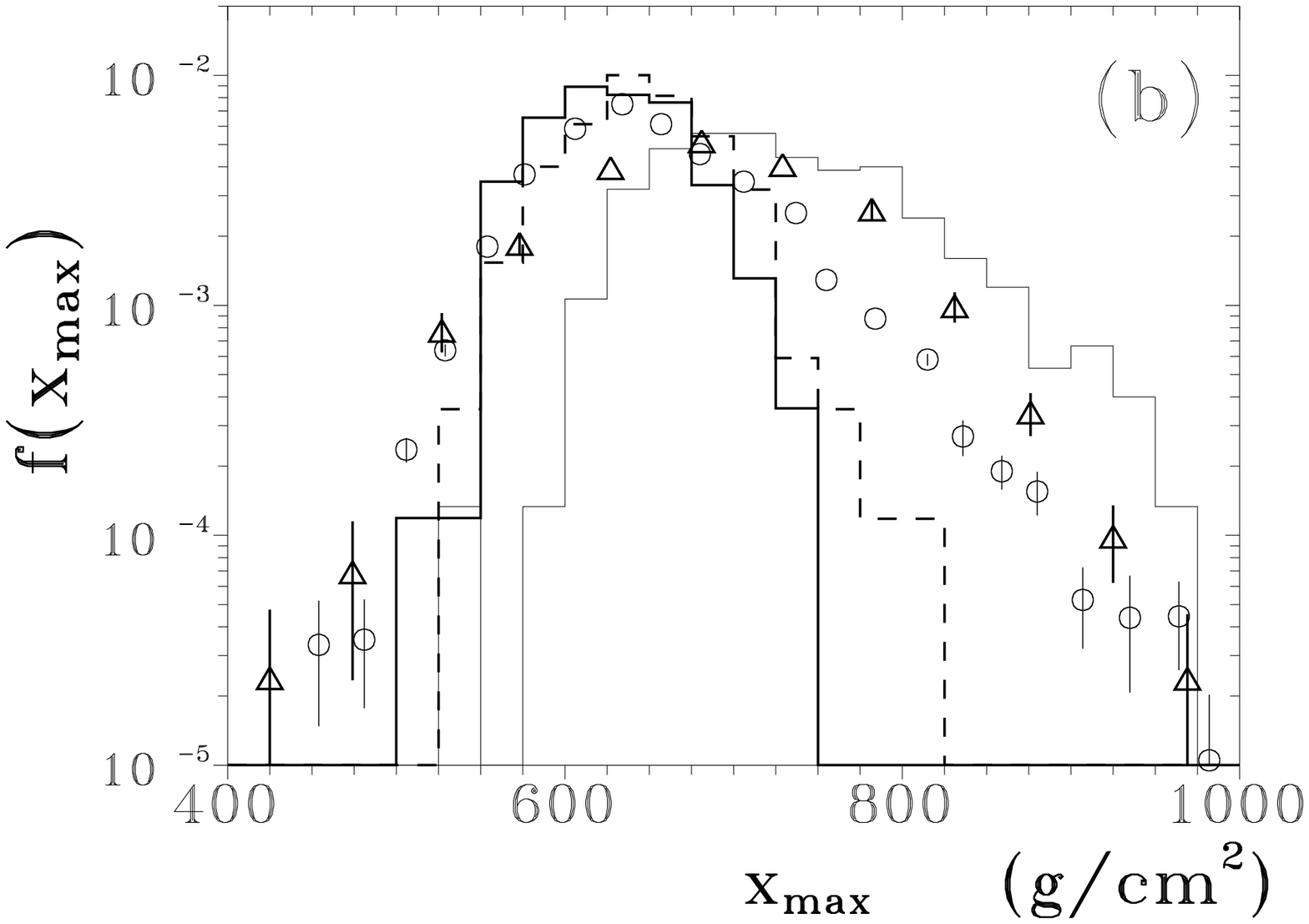,width=9.5cm}
 \psfig{file=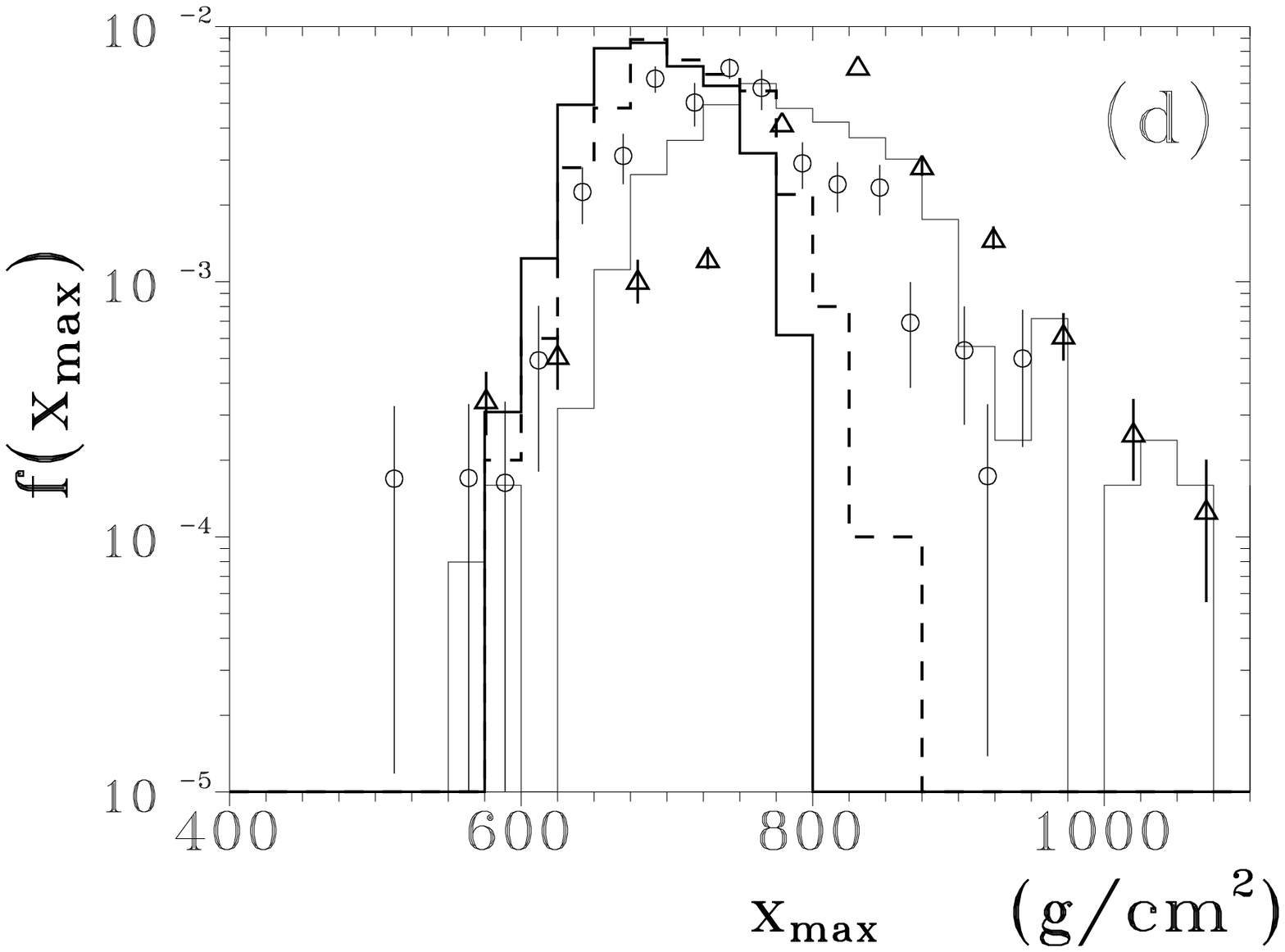,width=9.5cm}}
\caption{
Distributions of depths of shower maximum
calculated using the GMC model for showers of energies from
10$^{17}$ eV to 3$\times $10$^{17}$ eV (a),
3$\times $10$^{17}$ eV to 10$^{18}$ eV (b),
10$^{18}$ eV to 3$\times $10$^{18}$ eV (c), and
3$\times $10$^{18}$ eV to 10$^{19}$ eV (d),
respectively,
for protons (thin histogram) and iron (thick histograms).
For the iron primaries the dashed line represents the model
without second step cascading process while
the solid one is obtained with this process taken into account.
The circles represents the Fly's Eye data from [21].
The Yakutsk data (triangles) from Ref.[22] for showers with average energy of
5$\times $10$^{17}$ eV and
5$\times $10$^{18}$ eV
are given in (b) and (d).}
\label{xmax2}
\end{figure}

Even with the ``thinning'' procedures simulations of showers above
$\sim 10^{17}$ eV are very time consuming, so the statistics of simulated
showers is always not as large as one wants it to be. But due to the other
possible methodical uncertainties (mainly concerning interaction models),
the increase of statistical accuracy is not a main problem here.
(For the present analysis about 100 showers were generated
for each energy to draw lines in Fig.\ref{xmax} and
200 shower samples were generated
for each distribution of $x _{\rm max}$ given in Fig.\ref{xmax2}).

The longitudinal EAS profile for given primary particle mass and energy
fluctuates according to random character of the collision process.
For large showers, the depth of the shower maximum can be determined
in individual cases.
Simulated longitudinal shower development curve was sampled each 20
g~cm$^{-2}$ along the shower axis and the depth of the maximum
$x_{\rm max}$ was calculated using the polynomial interpolation
between these points (in log---lin scale) around the highest one
for each individual shower and then averaged over the whole shower sample.
It has to be stressed here that
our Monte Carlo results were not corrected for the
Fly's Eye detector effects
described above \cite{GS93}.

The $x_{\rm max}$ calculations were performed for primary protons and
iron nuclei. For the second case, the model with and without second
step cascading mechanism was used. Switching off the second step cascading
means that all primary wounded nucleons hodronize without any possibility
to collide with other nucleons (wounded or not wounded).
Results together with Fly's Eye and
Yakutsk data are given in Fig.~\ref{xmax}.

For the primary energy of about 10$^{15}$ eV our results for the GMC model
can be compared with predictions of other models
included in the CORSIKA program given in \cite{compa}. Respective values of
$x_{\rm max}$ and elongation rates
are presented in the Table~\ref{tab2} together with experimental results.
The Fly's Eye do not measure showers of energy about 10$^{15}$eV. In the table
there are shown two recently obtained results of DICE \cite{28} and HEGRA
\cite{27} experiments.
The technique used by DICE group is similar to the Fly's Eye geometrical
reconstruction of the longitudinal shower profile,
but using the Cherenkov photons, while the HEGRA use
the lateral Cherenkov light distribution to evaluate position of the shower
maximum.

Values of $x_{\rm max}$  and elongation rate for the first five models
were calculated in \cite{compa}
fitting the average longitudinal development profile
tabulated in 100 g~cm$^{-2}$ steps,
while our calculations
for GMC (sixth and seventh rows in Table~\ref{tab2})
were performed using the individual
shower profiles (like it is done in the discussed experiments). The difference
between the mean $x_{\rm max}$ and the depth of maximum of averaged
shower development curve is expected due to the asymmetry of the
shower profile and its large fluctuations.
It is seen when comparing
sixth and eighth rows in Table~\ref{tab2} where for the same simulated showers
both averaging methods were applied.
Taking this into account we can conclude
that the GMC interaction model at about 10$^{15}$ eV does not differ much
from other event generators commonly used in cosmic ray simulations.

Concerning the very high energies,
GMC results for proton primaries are close to this predicted in
\cite{kal95} (solid lines in Fig.~\ref{xmaxexp}($b$)).
Also the elongation rate is much higher than the one obtained
in \cite{GS93} and used in \cite{bird93} (presented in Fig.~\ref{xmaxexp}($a$)).
For iron primaries the situation is different.
Average depths of EAS maxima in the GMC model (also without second step
cascading) are much deeper
in the atmosphere then all the theoretical predictions given
in Fig.~\ref{xmaxexp}($b$). At the energy of 10$^{17}$ eV our result
is similar to the ``pure Monte Carlo'' (without 25 g~cm$^{-2}$ correction)
of \cite{GS93} for the minijet model (however, the elongation rate is
quite different).

The role of the second step cascading process is clearly seen in
Fig.~\ref{xmax}. It increases slowly with energy giving the positions
of the shower maxima higher in the atmosphere of about 10 g~cm$^{-2}$
at 10$^{16}$ eV and 30 g~cm$^{-2}$ at 10$^{19}$ eV.
Thus, if without this process the experimental data show the
``heavier than iron'' mass of primary cosmic rays, then with the second step
cascading taken into account data points lay between the proton
and iron mass predictions. The elongation rate changes from 76 g$\:$
cm$^{-2}$
/decade
to 70 g~cm$^{-2}$
/decade
 when the second step cascading is switched on for
pure iron mass (for protons is equal to 63 g~cm$^{-2}$
/decade). All these
values are in good agreement with the measured rates as can be seen in
Fig.~\ref{xmax}.

This allows us to closer investigate EAS maxima behaviour
looking for a possibility of primary mass determination.
Fig.~\ref{xmax} may suggests that the high-energy cosmic ray flux is
pure iron, but
the average value of $x_{\rm max}$ is one of parameters describing
the shower longitudinal development.
It should be remembered that the $x_{\rm max}$ is roughly
proportional to logarithm of primary mass and taking into account the
possible systematics discussed above its mean value is not the best
observable to determine the primary cosmic mass spectrum.
As has been said,
the development curves
fluctuate according to the probabilistic nature is the interaction
processes. The range of these fluctuations gives additional
information about the nature of cosmic ray primary particle. It is
expected that, not only the mean depth of shower maximum is smaller
for heavy primaries, but also the width of $x_{\rm max}$ distribution
has to be smaller. In Fig.~\ref{xmax2} $x_{\rm max}$ distributions
measured and calculated for pure proton and iron spectra are given.

Two different and important features can be found in these distributions.
One is an overall width related to the primary cosmic ray mass spectrum
and the second is the decrement or falling slope
of $x_{\rm max}$ distributions related to the proton--air interaction
cross section. If this slope is characterized by an exponential, the
exponential slope $\lambda$ can be identified as proton
interaction length.
The consistency between measured slopes and cross section values
used in the CORSIKA code is seen. The measured slope ($\lambda_{p-air}$)
can be approximated as 69, 58 and 67 g~cm$^{-2}$ from data in Figs.\ref{xmax2}
($a$), ($b$) and ($c$), respectively, while the interaction length from
simulated pure proton spectra is 64, 70 and 61 g cm$^{-2}$.
According to the statistical significance of the measurement (approximated
error is about of $\pm$7 g cm$^{-2}$) and simulations the agreement is
very good.

The falling part of $x_{\rm max}$ distribution is determined by the
proton (or, in general, light) cosmic ray primaries, while the
initial one is mainly generated by the showers induced by heavy
particles.
Calculations for pure iron spectrum without taking into
account the second step cascading process (dashed histograms)
clearly fail to reproduce the shape of
$x_{\rm max}$ distribution at small
$x_{\rm max}$ values. The predicted maxima are too deep in the
atmosphere. This is seen also in Fig.~\ref{xmax}.
The second step cascading shifts the maxima higher
(thick solid histograms) just to the altitude needed to explain the
initial part of $x_{\rm max}$ distribution. For the deeper developed showers,
however, there have to be a significant amount of lighter primaries in
the cosmic ray flux.
Because the slope of decrement of $x_{\rm max}$ distribution is
very close to the one expected for protons we can assume that there
is a mixture of iron, protons, and particles of an intermediate mass
(if needed) in the primary particle mass spectrum.
From the performed calculations we are able to try to determine the fraction
of iron and protons in primary cosmic ray spectrum in each of four
energy regions displayed in Fig.~\ref{xmax2}. The estimation of a proton
fraction has been performed
by comparing the large $x_{\rm max}$ part of the distribution obtained for pure
proton spectrum with respective measured frequencies. The iron abundance
was estimated by comparing initial parts of respective $x_{\rm max}$
distributions.
Results of such estimation are given in Table~\ref{tab} in the
comparison with values frequently used in the literature
given in \cite{KK}.

The abundances of He, CNO, and Ne--S groups are partially included
to estimated by us proton and iron fluxes which allows us to give
a strong statement only about the upper limits of proton and iron group
fluxes. Additionally because of $\log (A)$ dependence
of the shower longitudinal profile on the primary particle mass
(Ne--S showers are very close to iron showers) the correspondence between
our results with other presented in Table~\ref{tab} should be examined
with care. However, for the comparison, we have calculated mean values
of $x_{\rm max}$ for compositions listed in Table~\ref{tab} for
three energies of interest. They are given in the Table~\ref{tab3}.

The accuracy of the mass composition estimated using results
presented in Fig.~\ref{xmax2} is certainly not better than about 10\% mainly
due to small statistics (of the respective $x_{\rm max}$
distribution wings) of simulated shower samples.
More accurate estimations could be, however, questionable bearing in mind
the uncertainties concerning details of the interaction model
and the possible systematics related to detector effects (second error in
Fly's Eye data reported in Table~\ref{tab3} \cite{GS93}) not included in our
calculations.

As is seen in the Table~\ref{tab}, no significant change in the
cosmic ray chemical composition is found. At this point our
conclusions are similar to the one of the \cite{diakoncalga,kal95,KK}.
Such change would support the attractive interpretation of the change
of the character of
cosmic rays seen at about 10$^{18}$ eV (change of the energy spectrum index)
as the manifestation of appearing of the second component (non-diffusive or
extragalactic) of cosmic ray flux. However, according to calculations
presented in this paper it could hardly be a pure proton one and
even the ``normal'' composition seems to be ruled out.
We leave open the question of whether it could be as predicted by the
diffusion model.
Differences between measurements and calculations are still within
the possible systematics (see \cite{GS93}) as is shown in Table~\ref{tab3}.
The significant amount of heavy and very heavy nuclei is still seen at about
10$^{19}$ eV.

\section{Conclusions}

We have shown that extensive air showers initiated by ultrahigh-energy
cosmic rays can be described using the Monte Carlo
simulation code CORSIKA with the geometrical multichain model for
nucleus--nucleus interactions.
Data from Fly's Eye and Yakutsk experiments do not contradict significantly
the cosmic ray mass composition obtained using
Peters--Zatsepin diffusion model enriched by heavy nuclei at energy of
about 10$^{17}$ eV.
However the restoration
of the normal composition at 10$^{19}$ eV predicted by the diffusion
model seems to be unsuitable.
The lightening of primary cosmic ray mass spectrum above 10$^{18}$ eV
looks to be much weaker than reported by Fly's Eye group.

\begin{figure}
 \centerline{
 \psfig{file=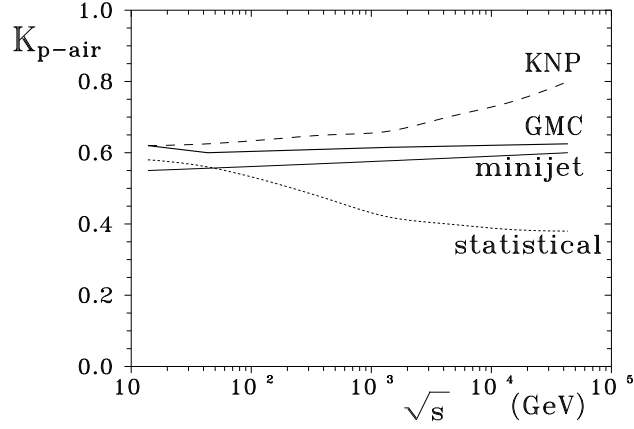,width=9.5cm}}
\caption{Inelasticity coefficient of $p$--$air$ collisions as a function
of interaction energy.
Three models of $p$--$air$ used in Ref.[12] analysis are presented
(thin lines) in comparison with the GMC model predictions (thick line).}
\label{kfenppa}
\end{figure}

It is important to note that this conclusion
is obtained using the interaction model which does not
introduce any ``extraordinary'' effects which starts to dominate
the interaction picture at ultrahigh energies. Our model for
proton--proton inelastic collision is, at this point, a very
``conservative'' one [see, e.g., Fig.~\ref{kfenppa}].
The agreement with data is mainly a consequence
of the second step cascading process which has to be present
in high-energy nucleus--nucleus interactions.
This mechanism leads to the significant increase on the
inelasticity for nuclei collisions without changing the hadron--hadron
interaction characteristics.

\begin{table}
\begin{center}
\caption{Depths of shower maximum at $10^{15}$ eV (g cm$^{-2}$)
and elongation rate (g cm$^{-2}$/decade)
for different models incorporated the CORSIKA program [Ref.(20)]
(first five rows) and the GMC model in comparison with some measured
values.}
\begin {tabular}{|c|cc|cc|}
\hline
\ \ \ Primary particle \ \ \  & \ &proton\ \ \ \ \  \
& \ &iron \  \ \ \ \ \ \  \ \\
\ & $x_{\rm max}$ &elongation rate&$x_{\rm max}$ &elongation rate \\
\hline
VENUS
& 574 & 71 & 439 & 84 \\
QGSJET
& 576 & 72 & 442 & 87 \\
SIBYLL
& 592 & 73 & 458 & 96 \\
HDPM
& 599 & 78 & 444 & 91 \\
DPMJET
& 560 & 68 & 428 & 75 \\
GMC
\tablenote{with second step cascading;
individual shower longitudinal profile.}
& 600 & 63 & 483 & 70 \\
GMC
\tablenote{without second step cascading;
individual shower longitudinal profile.}
      & 600 & 63 & 489 & 76 \\
GMC\tablenote{average longitudinal shower profile
(with second step cascading).}
      & 561 & 68 & 442 & 74 \\
\hline
\hline
%\end{tabular}
%\begin {tabular}{|c|cccc|}
experiment & x$_{\rm max}$ & \ & \ & \ \\
\hline
DICE
\tablenote{Ref.(27)} & 540 & \ & \ & \ \\
HEGRA
\tablenote{Ref.(28)} & 480 $\pm$ 30 & \ & \ & \ \\
\hline
\end{tabular}
\label{tab2}
\end{center}
\end{table}

\begin{table}
\begin{center}
\caption{Relative abundances of nuclei groups.}
\begin {tabular}{|r|ccccc|}
\hline
Group of nuclei & protons & He & CNO & Ne--S & Fe \\
\hline
present work\ \ \  \ \ \ \ \ \ \ \ \ \ \  \ \ \ & \ & \   & \ & \ &\  \\
10$^{17}\div 3 \times  10^{17}$ eV &10\%&\ &\ &\ & 50\% \\
$3 \times 10^{17}\div 10^{18}$ eV &20\%&\ &\ &\ & 60\% \\
10$^{18}\div 3 \times   10^{18}$ eV &30\%&\ &\ &\ & 60\% \\
$3 \times   10^{18}\div 10^{19}$ eV &30\%&\ &\ &\ & 50\% \\
\hline
Normal composition\ \ \  \ \ \ \ \ & \ & \ &\ &\ &\ \\
\ & 32\% & 23\% & 21\% & 14\% &10\% \\
Heavy composition [24]\ \ \ \ & \ & \ &\ &\ &\ \\
\ & 14\% & 23\% & 26\% & 13\% & 24\% \\
Diffusion model [25]\  \ \ \ \ \ \ \ & \ & \ &\ &\ &\  \\
10$^{16}$ eV & 24\% &24\% & 24\% &16\% &12\% \\
10$^{17}$ eV & 14\% &14\%& 25\%& 24\%& 23\% \\
Stanev {\it et al.} [26] \ \ \ \ \ \ \ \ \ \ \ & \ & \ &\ &\ &\ \\
10$^{15}$ eV & 7\% &31\% & 18\% &31\% &13\% \\
10$^{17}$ eV & 8\% &17\%& 17\%& 37\%& 21\% \\
\hline
\end{tabular}
\label{tab}
\end{center}
\end{table}

\begin{table}
\begin{center}
\caption{Mean value of $x_{\rm max}$ (g cm$^{-2}$)
calculated for CR compositions given in
Table II compared with Fly's Eye measured values [14]}
\begin {tabular}{|c|cccc|c|}
\hline
\  &\  &\ &  composition \ \ \ \ \ \ \ \
 &\ & \  \\
\ \ \ Energy\ \ \  & Normal & Heavy & Diffusion &  Stanev {\it et al.}\ \ \  &
\ \ \ Fly's Eye result \ \ \ \\
\hline
10$^{17}$eV & \ \ \ \ \ \ 670\ \ \ \ \ \  &\ \ \ \ \ \  650\ \ \ \ \ \  &
\ \ \ \ \ \  645\ \ \ \ \ \
&\ \ \ \ \ \  640\ \ \  \ \ \  &\ \ \ \ \ 625$\pm$5$\pm$20 \ \ \ \ \\
10$^{18}$eV & 735 & 720 & 715 & 710 & \ \ \ \ \ 683$\pm$3$\pm$20 \ \ \ \ \\
10$^{19}$eV & 805 & 785 & 780 & 780 & \ \ \ \ \ 760$\pm$10$\pm$20
\tablenote{E$>3 \: 10^{18}$eV}\ \ \\
\hline
\end{tabular}
\label{tab3}
\end{center}
\end{table}

\end{document}